\documentclass[a4paper,11pt]{article}
\usepackage{pos}
\usepackage{cleveref}
\newcommand{\aff}[2]{\noindent\textsuperscript{#1}#2\par}

\title{The TAMBO Project and the TAMBITO Prototype}

\author{Pablo Fernández-Menéndez* for the TAMBO Collaboration}

\affiliation{Universidade de Santiago de Compostela, Instituto Galego de F\'isica de Altas Enerx\'ias (IGFAE), R\'ua de Xoaqu\'in D\'iaz de R\'abago, s/n, Santiago de Compostela, 15705, Spain}
\emailAdd{pablo.fer@cern.ch}

\abstract{
The detection of astrophysical PeV-scale neutrinos by IceCube and KM3NeT observatories has opened the window to probe new phenomena in the universe. However, a deeper study of the sources remains limited by the small astrophysical flux and the sizable atmospheric neutrino background. The Tau Air-Shower Mountain-Based Observatory (TAMBO) aims to overcome these limitations by detecting tau neutrinos from 1\,PeV to 1\,000\,PeV, an energy range in which backgrounds are largely suppressed and which lies between the reach of optical-Cherenkov observatories and that of radio-based experiments. TAMBO will employ up to 5,000 particle detectors installed on the face of a steep hill in the Colca Valley (Perú) to detect tau neutrinos interacting with the face of the canyon in front via the decay products of the associated tau leptons.
In this contribution, we will describe the current status of the project and its R\&D, focusing on the 100-unit array TAMBITO prototype, the projected timeline, and the physics potential that this unique detector brings to the field.

}

\FullConference{29th European Cosmic Ray Symposium (ECRS2026)\\
29 June - 3 July 2026\\
Santiago de Compostela, Spain\\}

\begin{document}
\maketitle

\section{Introduction} \label{intro}

The discovery of a diffuse astrophysical neutrino flux by IceCube in 2013~\cite{IC2013} opened a new window onto some of the most energetic phenomena in the universe: the field of high-energy neutrino astronomy. Since then, IceCube has found evidence for neutrino emission from the blazar TXS~0506+056, from the active galaxy NGC~1068 and from the Galactic plane~\cite{IC_TXS, IC_NGC, IC_Gal}, and KM3NeT has reported an event with an estimated energy of about 220\,PeV~\cite{KM3}.

At the highest energies, ARIANNA~\cite{ARIANNA}, ARA~\cite{ARA} and the Pierre Auger Observatory~\cite{Auger_nu} have set upper limits on the neutrino flux, and a variety of experiments have been proposed to study astrophysical neutrinos further. Existing and future detectors fall into two main technological classes. The first comprises optical-Cherenkov detectors, which instrument large volumes of water or ice with photosensors to capture Cherenkov light; examples are IceCube-Gen2~\cite{IceCubeGen2}, KM3NeT/ARCA~\cite{ARCA} and P-ONE~\cite{PONE}, which are sensitive from the TeV scale up to a few PeV. The second comprises radio-based detectors, which detect radio signals from neutrino-induced cascades or air showers with antennas and are sensitive to ultra-high energies, from about 10\,PeV, with most of their sensitivity at EeV energies; examples are HERON~\cite{HERON}, GRAND~\cite{GRAND}, RNO-G~\cite{RNOG} and the Radar Echo Telescope (RET)~\cite{RET}.

However, both point-source searches and measurements of the spectrum above 1~PeV suffer from large backgrounds, dominated by atmospheric neutrinos and several orders of magnitude above the signal~\cite{TAMBO_arXiv}. Consequently, and to avoid a trials-factor penalty, experiments are forced to restrict point-source searches to X-ray or $\gamma$-ray source catalogs.

The Tau Air-Shower Mountain-Based Observatory (TAMBO) is designed to address both issues, offering an almost background-free sample of tau neutrinos, improved direction reconstruction, and sensitivity that bridges the high-energy and ultra-high-energy regimes~\cite{TAMBO_arXiv,TAMBO_NA}.

\section{The TAMBO Detector} \label{detector}

\begin{figure}[h!]
    \centering
    \includegraphics[width=0.8\linewidth]{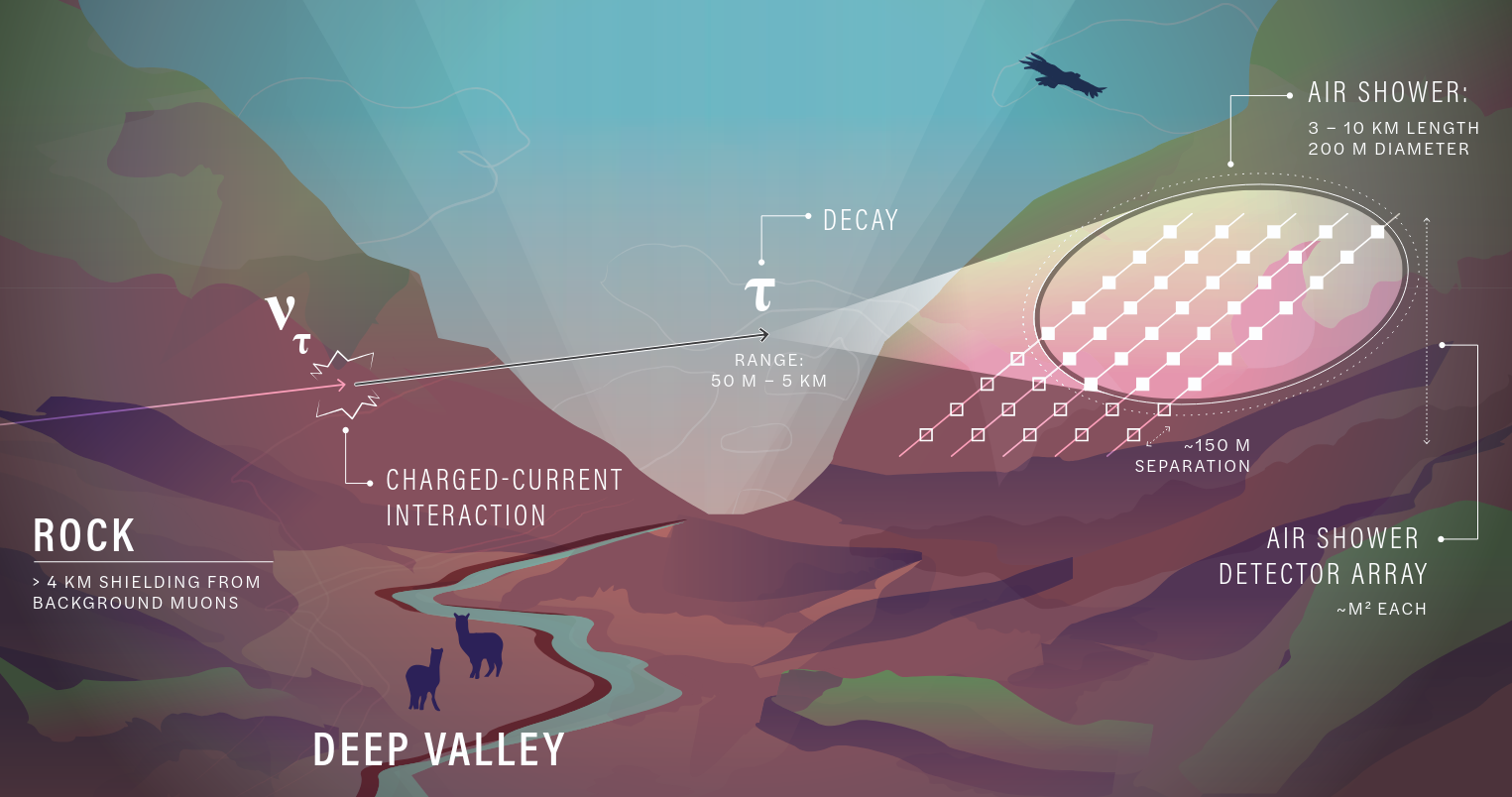}
    \caption{\textit{Infographic of the TAMBO detection principle: a tau neutrino interacts in the rock on one side of the valley, the resulting tau lepton exits into the air and decays, and the air shower is observed by the detector array on the opposite face.}}
    \label{fig:tambo_add}
\end{figure}

TAMBO exploits the topography of a deep valley to detect Earth-skimming high-energy tau neutrinos, as illustrated in \Cref{fig:tambo_add}. The idea of observing the showers of Earth-skimming taus from a mountain across a valley goes back to Ref.~\cite{Fargion} and was developed for the Andean deep-valley geometry in Ref.~\cite{Romero}. High-energy neutrinos can travel up to hundreds of kilometres through the Earth's crust, with an energy-increasing probability of interacting with the nuclei of the rock. In a tau-neutrino charged-current interaction, a short-lived charged tau lepton ($\tau$) is produced. In the 1--100\,PeV range, about 80\% of the neutrino energy is transferred to the $\tau$, which travels between 50\,m and 5\,km before decaying~\cite{TAMBO_ESPP}.

When the interaction occurs close to the surface of the valley face, the $\tau$ can exit into the air and decay, and the resulting shower crosses the valley and reaches the hill in front. The detector array will thus observe tau neutrinos arriving just below the horizon, through the $\tau$ decay modes that can produce a detectable shower: the hadronic and electronic modes, which together account for about 82\% of decays~\cite{TAMBO_ICRC23, TAMBO_ESPP}. Conversely, $\nu_e$ charged-current interactions produce electrons that lose their energy almost immediately in the rock, and $\nu_\mu$-produced muons rarely convert into an air shower. At these energies, the cosmic $\nu_\tau$ signal also dominates over the atmospheric $\nu_\tau$ background, so any event is likely to be of cosmic origin~\cite{TAMBO_arXiv}.

The characteristic length the $\tau$-induced shower needs to develop is between 3 and 10\,km, which added to the over 4\,km of rock needed to filter most cosmic muons, constrain candidate site locations for the detector in the valley. The Colca Valley in the Peruvian Andes offers the required topography at several candidate sites; the selection of the optimal site is ongoing, including a responsible-siting programme carried out together with local stakeholders~\cite{Swart}. From the Colca Valley, the detector would have the Galactic Centre and much of the Galactic plane in its field of view and, thanks to the Earth's rotation, would cover nearly the full sky~\cite{TAMBO_ESPP}.



\begin{figure}[h!]
    \centering
    \includegraphics[width=0.55\linewidth]{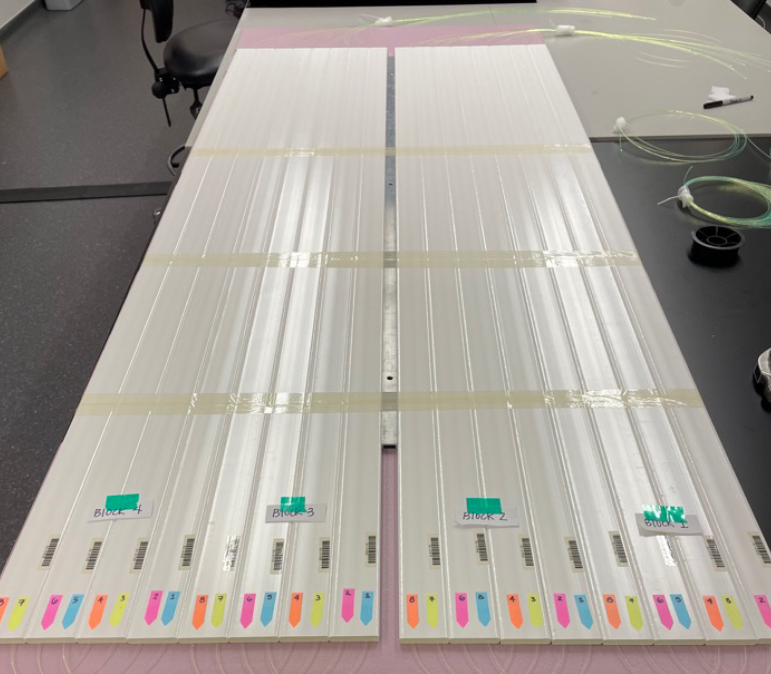}
    \caption{\textit{Image of an assembled plastic-scintillator panel for TAMBO~\cite{TAMBO_ICRC23}.}}
    \label{panels}
\end{figure}

In its nominal configuration, TAMBO comprises 5\,000 detection units with a cross-section of order 1\,m$^2$, spaced 150\,m apart on a triangular grid on the valley slope and covering about 50\,km$^2$~\cite{TAMBO_arXiv,Zhelnin_LP}. The reference detection units are plastic-scintillator panels, whose signal is guided by wavelength-shifting fibers and read out by silicon photomultipliers (SiPMs), a cost-effective option that is easy to deploy (see \Cref{panels}).

Water-Cherenkov tanks are a complementary option for the detection units. They offer a larger angular acceptance, sensitivity to shower photons and even the possibility of discriminating between electrons and muons~\cite{TAMBO_ICRC23}. However, the logistics of transport, installation and operation remain challenging and are the subject of extensive R\&D.

\section{Data Acquisition and Expected Performance}

\begin{figure}
    \centering
    \includegraphics[width=0.6\linewidth]{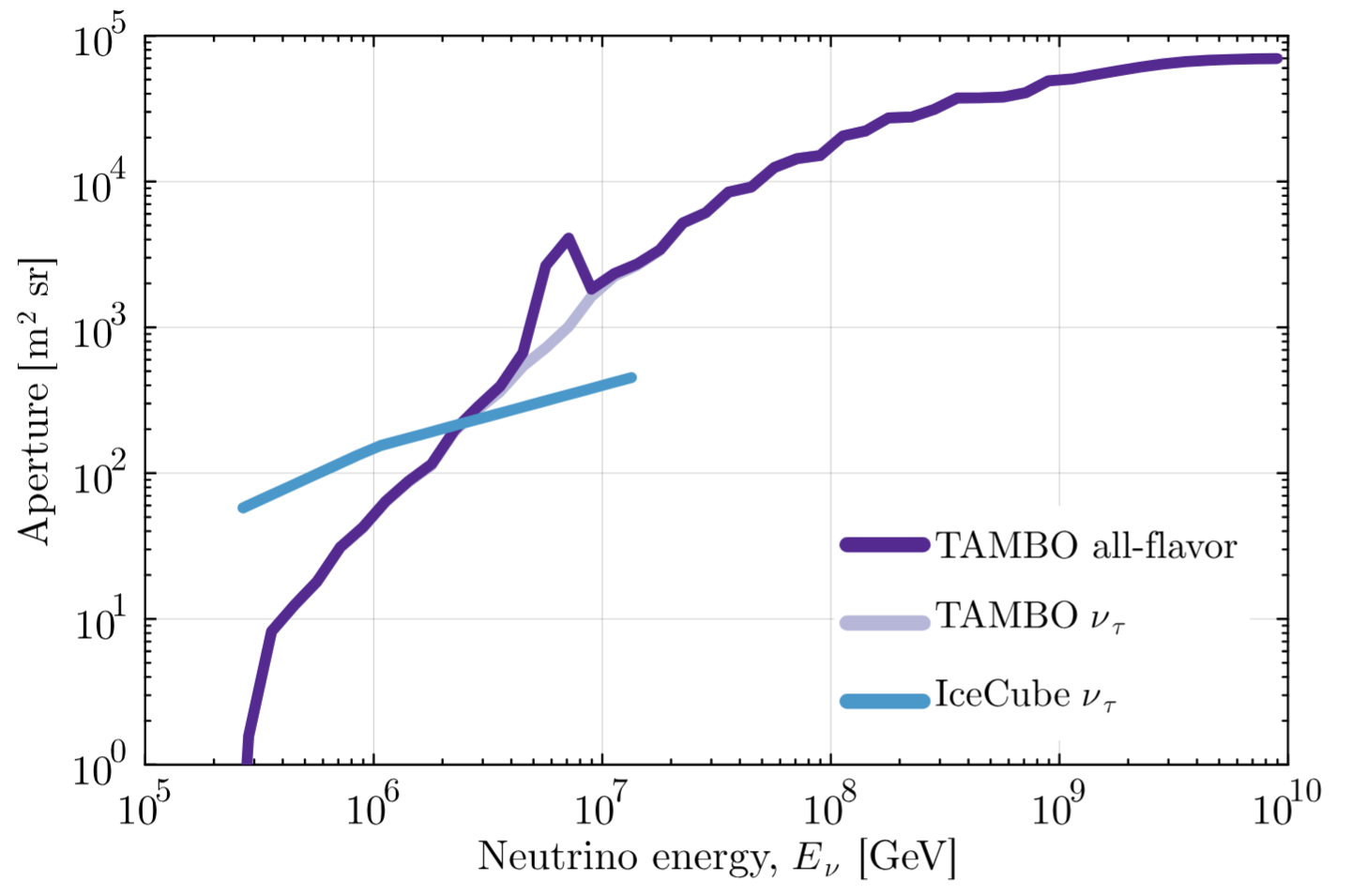}
    \caption{\textit{Expected acceptance (aperture) of TAMBO, estimated from the simulation of an array of 5\,000 detection units
spaced 150\,m apart at a candidate site in the Colca Canyon. The corresponding IceCube acceptance
is shown for comparison~\cite{TAMBO_arXiv}.}}
    \label{fig:acceptance}
\end{figure}

For 5\,000 detection units with 150\,m spacing in a 4.5\,km wide valley, the aperture of TAMBO grows with the energy of the incoming tau neutrino, approximately as $E_\nu^{1.5}$. It overtakes that of IceCube at around 2\,PeV and flattens above 1\,EeV, where the $\tau$ decay length exceeds the width of the valley (\Cref{fig:acceptance})~\cite{TAMBO_arXiv,Lazar_ICRC25}.

A large fraction of the shower particles leave a signal in the plastic-scintillator panels, which record the arrival time and charge from which the energy and direction of the initial neutrino are inferred. In the nominal design, the trigger suppresses random coincidences and cosmic-muon backgrounds by requiring coincident signals in at least three neighbouring units together with a total signal above the equivalent of 30 particles~\cite{TAMBO_ESPP}. Combined with direction-based selection cuts, this keeps the expected background to $\lesssim 1$ event in 10 years~\cite{TAMBO_ESPP}.

\begin{figure}
    \centering
    \includegraphics[width=0.8\linewidth]{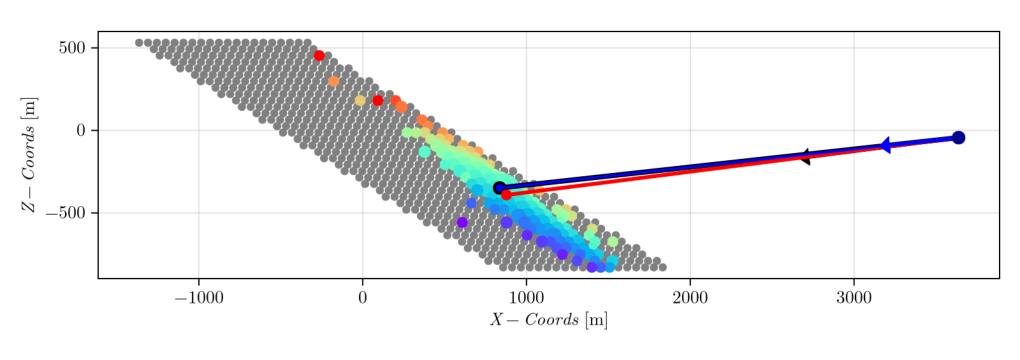}
    \caption{\textit{True and direction reconstruction tau-neutrino-induced shower in TAMBO~\cite{Zhelnin_LP}.}}
    \label{fig:reco}
\end{figure}

Simulations show that, given the geometry of the valley and the detector
layout, a timing resolution of $\lesssim 7$\,ns allows the neutrino direction
to be reconstructed with a precision of $0.5^\circ$~\cite{Zhelnin_LP}.
A sub-degree pointing capability is crucial for point-source searches.
An example is shown in \Cref{fig:reco}.
 
\section{Physics Goals and Potential} \label{physics}

Thanks to its low background and high-precision direction reconstruction, TAMBO can overcome the current limitation of point-source searches to X-ray and $\gamma$-ray source catalogs. Each event detected by TAMBO is expected to be, with high confidence, of cosmic origin and therefore points to either a cosmic neutrino source or the cosmogenic neutrino flux. TAMBO will thus be able to deliver a neutrino-only source catalog and guide larger-statistics observatories in studying the identified sources further.

\begin{figure}
    \centering
    \includegraphics[width=\linewidth]{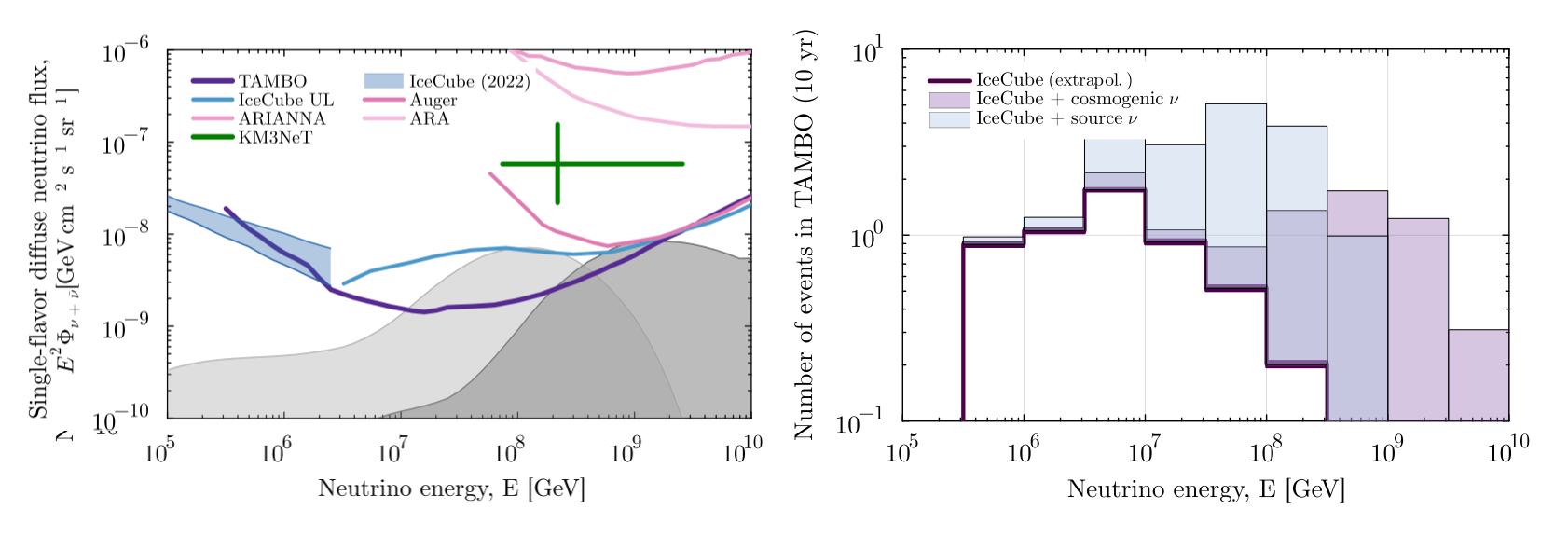}
    \caption{\textit{Left: upper limits at 90\% confidence level on the single-flavour diffuse astrophysical neutrino flux from various experiments, together with the projected TAMBO sensitivity; shaded areas correspond to neutrino-source and cosmogenic models. Right: expected number of neutrino events in 10 years of TAMBO data taking, assuming the astrophysical diffuse flux extrapolated from IceCube data plus additional neutrino-source and cosmogenic contributions~\cite{TAMBO_arXiv}.}}
    \label{fig:astro_flux}
\end{figure}

Regarding the diffuse astrophysical flux, TAMBO's projected sensitivity exceeds that of present observatories between about 1\,PeV and 1\,EeV, acting as a bridge between optical-Cherenkov and radio-based observatories, as shown in \Cref{fig:astro_flux}. Assuming only the IceCube-measured flux~\cite{Naab}, TAMBO expects on average 6 extragalactic neutrinos in ten years; with optimistic source or cosmogenic models~\cite{Rodrigues}, this number exceeds 20. Despite the small size of its data sample, the strength of TAMBO lies in the purity of the data. In addition, by efficiently tagging tau neutrinos and in combination with all-flavour data, TAMBO will improve the measurement of the flavor composition of this flux~\cite{Snowmass}.

Besides the astrophysical neutrino programme, TAMBO will be able to probe phenomena beyond the Standard Model (SM). The tau neutrino is the least-studied SM particle: IceCube has reported only a handful of tau-neutrino candidates~\cite{ICtau1,ICtau2}, and neutrino--nucleon cross-section measurements at PeV energies and above~\cite{ICxs} remain largely unexplored tests of the SM. TAMBO will also be able to search for neutrino--dark-matter interactions, sterile neutrinos, pseudo-Dirac mechanisms, Lorentz violation and quantum-gravity effects at PeV to EeV energies~\cite{TAMBO_arXiv}.

Finally, the valley detector layout enables simultaneous measurements of the electromagnetic and muonic components of cosmic-ray air showers, which could provide additional information on the muon puzzle, the deficit of muons predicted by hadronic interaction models relative to the measured numbers~\cite{Albrecht}.

\section{R\&D Phase, TAMBITO and Project Timeline} \label{tambito}

The TAMBITO prototype is currently in the component-construction phase and comprises the R\&D for 100 plastic-scintillator units to be installed in the Colca Valley region~\cite{Queens}. Its goals are to test and validate deployment techniques on real terrain, to check calibration, data acquisition, timing and background rejection, to perform cosmic-ray measurements, and to introduce the project to the local community and develop it hand in hand with them. TAMBITO follows the nominal design for the plastic-scintillator detection units. It is part of a 2025--2028 programme, with deployment and operation expected during 2027 and 2028~\cite{TAMBO_ESPP}.

Following the R\&D programme, TAMBO is expected to be deployed in stages, with a 1\,500-module array operating from 2029, followed by two further deployments of 1\,500 and 2\,000 modules to reach the 5\,000-module design~\cite{TAMBO_ESPP}. This schedule remains subject to funding, site selection and the results of the prototype.

\section{Conclusions} \label{conclusions}

TAMBO is a planned deep-valley tau-neutrino telescope that uses mountain topography as a shield and as a target, with
5\,000 scintillator-based units observing the air showers of decaying taus exiting the rock of the other side of the valley. It bridges the energy gap between Cherenkov and radio detectors, and offers a
nearly background-free astrophysical sample with sub-degree pointing. This
enables measurements of the diffuse flux between 1 and 1\,000\,PeV and the
production of astrophysical neutrino-only source catalogs.

The TAMBITO prototype and associated R\&D are ongoing and will run until 2028, demonstrating the technology and the deployment procedures.

Additionally, the TAMBO project seeks to engage with local community to develop the project ethically and environmentally respectful \cite{Swart}.

\vspace{1cm}\emph{Tambos were buildings found along the extensive Inca roads typically contained supplies, served as lodging for itinerants. Tambos were spaced along Inca roads, generally about one day's travel apart.}

\begin{center}
\par\rule{0.6\textwidth}{0.5pt}
\end{center}

\begingroup\small
\section*{Full Author List: TAMBO Collaboration}
\noindent
Carlos Argüelles\textsuperscript{1}, Marvin V. Ascencio-Sosa\textsuperscript{2}, J. A. Balsells\textsuperscript{3}, José Bazo\textsuperscript{4}, Meir Behar\textsuperscript{5}, Miguel Bonnett del Alamo\textsuperscript{6}, Christopher Briceño\textsuperscript{7}, Mauricio Bustamante\textsuperscript{8}, Saneli Carbajal\textsuperscript{5}, Kiara Carloni\textsuperscript{1}, José Carpio Dumler\textsuperscript{8}, Cesar Castromonte\textsuperscript{9}, Víctor Centa\textsuperscript{4}, Thomas Chetaille\textsuperscript{1}, Luz Condori\textsuperscript{10}, Giulia F. Curatola Fernández\textsuperscript{11}, Jaco de Swart\textsuperscript{12}, Diyaselis Delgado\textsuperscript{1}, Deborah Delgado Pugley\textsuperscript{13}, Zlatan Dimitrov\textsuperscript{14}, Tommaso Dorigo\textsuperscript{15}, Bruno Díaz Arrascue\textsuperscript{4}, Erick Wilfredo Díaz Saborio\textsuperscript{3}, Erin Ewart\textsuperscript{16}, Anatoli Fedynitch\textsuperscript{17}, Pablo Fernández\textsuperscript{18}, Nishat Fiza\textsuperscript{4}, Alberto Gago\textsuperscript{4}, Gustavo Garcia Flores\textsuperscript{10}, Alfonso García\textsuperscript{19}, Carlo Gulizia\textsuperscript{20}, Zigfried Hampel-Arias\textsuperscript{21}, Hamza Hanif\textsuperscript{15}, Raul Henriquez-Ortiz\textsuperscript{22}, Connor Houston\textsuperscript{23}, Sebastián Huerta\textsuperscript{4}, K. Hughes\textsuperscript{24}, Jeffrey Hyde\textsuperscript{25}, Juan A. Jave-Sagastegui\textsuperscript{4}, Nicholas W. Kamp\textsuperscript{1}, Ali Kheirandish\textsuperscript{26}, Suhyeon Kim\textsuperscript{4}, Tanvi Krishnan\textsuperscript{1}, Jeffrey P. Lazar\textsuperscript{27}, Peter M. Lewis\textsuperscript{28}, Hermes León Vargas\textsuperscript{29}, Qinrui Liu\textsuperscript{30}, Daniel Menéndez\textsuperscript{4}, Marco Milla\textsuperscript{4}, Moises Nuñez\textsuperscript{7}, Omar Ortiz\textsuperscript{31}, Luis Otiniano\textsuperscript{32}, Josue Pareja Contreras\textsuperscript{10}, Alberto Peláez\textsuperscript{7}, Héctor Eduardo Pérez Figueroa\textsuperscript{3}, Oscar F. Rodriguez Rios\textsuperscript{4}, Enrique Rojas Villalba\textsuperscript{33}, Andres Romero-Wolf\textsuperscript{28}, José Rodrigo Sacahui Reyes\textsuperscript{3}, A. Salcedo Gomez\textsuperscript{24}, D.E. Scipion Castillo\textsuperscript{34}, S. Shefali\textsuperscript{1}, José Enrique Solano del Castillo\textsuperscript{35}, Carlos Javier Solano Salinas\textsuperscript{36}, Dennis Soldin\textsuperscript{23}, Luciano Stucchi\textsuperscript{5}, Damián Tapia Romero\textsuperscript{4}, Jimmy Tarrillo\textsuperscript{7}, William G. Thompson\textsuperscript{1}, Ricky Ticse\textsuperscript{4}, Luz E. Tiscareño Montoya\textsuperscript{4}, Jorge A. Torres\textsuperscript{23}, Stef Verpoest\textsuperscript{37}, Aaron C. Vincent\textsuperscript{38}, Pietro Vischia\textsuperscript{20}, Cynthia Vizcarra\textsuperscript{39}, Philip L. R. Weigel\textsuperscript{1}, Pablo Yanyachi\textsuperscript{40}, Jose Zavala\textsuperscript{7}, and Pavel Zhelnin\textsuperscript{1}
\medskip

\aff{1}{Department of Physics \& Laboratory for Particle Physics and Cosmology, Harvard University, Cambridge, MA 02138, USA}
\aff{2}{Institut für Experimentalphysik, Universität Hamburg, Luruper Chaussee 149, 22761 Hamburg, Germany}
\aff{3}{IFIM, Escuela de Ciencias Físicas y Matemáticas, Universidad de San Carlos de Guatemala, 01012, Ciudad de Guatemala, Guatemala}
\aff{4}{Sección Física, Departamento de Ciencias, Pontificia Universidad Católica del Perú, Apartado 1761, Lima, Perú}
\aff{5}{Universidad del Pacífico, Jr. Gral, Jirón Luis Sánchez Cerro 2141 Lima, Jesús María 15072, Peru}
\aff{6}{Universidad Nacional San Antonio de Abad del Cusco, Cusco, Perú}
\aff{7}{Electrical and Mechatronic Department, Universidad de Ingenieria y Tecnologia, Barranco 15063, Perú}
\aff{8}{Niels Bohr International Academy, Niels Bohr Institute, University of Copenhagen, DK-2100 Copenhagen, Denmark}
\aff{9}{Universidad Nacional de Ingeniería, Lima, Perú}
\aff{10}{Universidad Católica San Pablo, Arequipa, Perú}
\aff{11}{Institute of Geography, University of Bern, Hallerstrasse 12, 3012 Bern, Switzerland}
\aff{12}{Institute of Astronomy, University of Cambridge, Madingley Road, Cambridge CB3 0HA, United Kingdom}
\aff{13}{Departamento de Ciencias Sociales, Pontificia Universidad Católica del Perú, Apartado 1761, Lima, Perú}
\aff{14}{GATE Institute, Sofia University, 5 James Bourchier Blvd., 1164 Sofia, Bulgaria}
\aff{15}{INFN, Sezione di Padova, Italy, and Lulea University of Technology, Lulea, Sweden}
\aff{16}{Department of Physics, Engineering Physics and Astronomy, Queen's University, and Arthur B. McDonald Canadian Astroparticle Physics Research Institute, Kingston, Ontario, Canada}
\aff{17}{Institute of Physics, Academia Sinica, Taipei City, 11529, Taiwan}
\aff{18}{Instituto Galego de Física de Altas Enerxías (IGFAE), Universidade de Santiago de Compostela, 15782 Santiago de Compostela, Spain}
\aff{19}{Instituto de Física Corpuscular (IFIC), CSIC and Universitat de València, 46980 Paterna, València, Spain}
\aff{20}{Calle San Francisco 3, 33007 Oviedo, Principado de Asturias, Spain}
\aff{21}{Los Alamos National Laboratory, Los Alamos, NM 87545, USA}
\aff{22}{Facultad de Ciencias Naturales y Matemática, Universidad de El Salvador, Final 25 Avenida Norte, San Salvador, El Salvador}
\aff{23}{Department of Physics \& Astronomy, University of Utah, Salt Lake City, UT 84112, USA}
\aff{24}{Dept. of Physics, Center for Cosmology and AstroParticle Physics, Ohio State University, Columbus, OH 43210, USA}
\aff{25}{Department of Physics \& Astronomy, Swarthmore College, Swarthmore PA 19081 USA}
\aff{26}{Department of Physics \& Astronomy, University of Nevada, Las Vegas, NV 89154, USA}
\aff{27}{Department of Physics, Massachusetts Institute of Technology, Cambridge, MA 02139, USA}
\aff{28}{Department of Physics and Astronomy, University of Hawai'i at Mānoa, 2505 Correa Road, Honolulu, HI 96822, USA}
\aff{29}{Facultad de Ciencias, Universidad Nacional Autónoma de México, Ciudad Universitaria, 04510 Ciudad de México, México}
\aff{30}{Department of Physics, Simon Fraser University, Burnaby, BC V5A 1S6, Canada}
\aff{31}{Department of Physics, Universidad de Ingenieria y Tecnologia, Barranco 15063, Perú}
\aff{32}{Comisión Nacional de Investigación y Desarrollo Aeroespacial (CONIDA), Lima, 15046, Perú}
\aff{33}{MIT Haystack Observatory, 99 Millstone Road, Westford, MA 01886, USA}
\aff{34}{Instituto Geofísico del Perú, Calle Badajoz 169, Mayorazgo IV Etapa, Ate, Lima 15012, Peru}
\aff{35}{Department of Anthropology, University of Michigan, Ann Arbor, MI 48109, USA}
\aff{36}{Universidad Nacional Mayor de San Marcos, Lima, Perú}
\aff{37}{Department of Physics and Astronomy, University of Delaware, Newark, DE 19716, USA}
\aff{38}{Department of Physics, Engineering Physics and Astronomy, Queen's University, and Arthur B. McDonald Canadian Astroparticle Physics Research Institute, Kingston, Ontario, Canada, and Perimeter Institute for Theoretical Physics, Waterloo, Ontario, Canada}
\aff{39}{Universidad Continental, Arequipa, Perú}
\aff{40}{Universidad Nacional de San Agustín de Arequipa, Santa Catalina 117, Arequipa, Peru}
\endgroup

\end{document}